\documentclass[prc,aps,amssymb,showpacs]{revtex4}
\usepackage{graphicx}
\begin{document}

\title{Effect of Chiral Symmetry Restoration on Pentaquark $\Theta^+$ Mass and
Width\\ at Finite Temperature and Density}

\author{Xuguang Huang \footnote{e-mail:huangxg03@mails.tsinghua.edu.cn},
        Xuewen Hao \footnote{e-mail:haoxuewen@tsinghua.org.cn},
    and Pengfei Zhuang \footnote{e-mail:zhuangpf@mail.tsinghua.edu.cn}}
\affiliation{ Center of Theoretical Nuclear Physics, National Laboratory of Heavy Ion Collisions,
              Lanzhou 730000, China\\
              Physics Department, Tsinghua University, Beijing 100084, China}
\date{\today}

\begin{abstract}
We investigate the effect of chiral phase transition on the
pentaquark $\Theta^+$ mass and width at one-loop level of
$N\Theta^+K$ coupling at finite temperature and density. The
behavior of the mass, especially the width in hadronic medium is
dominated by the characteristics of chiral symmetry restoration at
high temperature and high density. The mass and width shifts of
positive-parity $\Theta^+$ are much larger than that of
negative-parity one, which may be helpful to determine the parity
of $\Theta^+$ in high energy nuclear collisions.
\end{abstract}

\maketitle

\section{Introduction}
Recently, an exotic baryon $\Theta^+$ was discovered firstly by
LEPS group at Spring-8 in the reaction $\gamma n\rightarrow
K^+K^-n$\cite{leps}, and was subsequently observed by many other
groups\cite{diana,clas,saphir,itep,clasnew,hermes,svd,cosy,yerevan,zeus,forzeus}.
It has $K^+ n$ quantum numbers(B=+1, Q=+1, S=+1), and its minimal
quark content must be $uudd \bar s$. The remarkable features of
the $\Theta^+$ are its small mass (1540MeV) and very narrow width
($<$25MeV)\cite{leps}. While the isospin of $\Theta^+$ is probably
zero\cite{saphir,clasnew,hermes,forzeus}, the other quantum
numbers including spin and parity have not been measured
experimentally yet. Theoretically, most of the works considers
$\Theta^+$ as a $J={1\over 2}$ particle because of its low
mass\cite{diak,jaffe,riska,zhang}, and some predict positive
parity\cite{diak,jaffe,riska,chiu,lipkin,hosaka,carl,liuyx,mathur}
and some suggest negative
parity\cite{zhang,mathur,lattice1,zhu,qsr,carlson,wu,song}.

The search for the pentaquark $\Theta^+$ is also extended to the
experiment of relativistic heavy ion collisions where the extreme
condition to form a new state of matter -- quark-gluon plasma
(QGP) can be reached. The STAR collaboration\cite{star1} reported
the progress of the pentaquark search in $p-p, d-Au$ and $Au-Au$
collisions at energy $\sqrt s = 200$ GeV, and the PHENIX
collaboration\cite{phenix} investigated the decay of the
anti-pentaquark $\bar\Theta^+ \rightarrow K^- \bar N$.
Theoretically, the baryon density modification\cite{navarra} on
the $\Theta^+$ mass was discussed with a phenomenological
density-dependent nucleon propagator\cite{kim}, and the $\Theta^+$
production in relativistic heavy ion collisions was studied in the
coalescence model\cite{chen}.

It is generally believed that there are two QCD phase transitions
in hot and dense nuclear matter. One of them is related to the
deconfinement process in moving from a hadron gas to QGP, and the
other one describes the transition from the chiral symmetry
breaking phase to the phase in which it is restored. From the QCD
lattice simulations\cite{lattice2}, the phase transitions are of
first order in high density region and may be only a crossover in
high temperature region. As the order parameter of chiral phase
transition, the dynamic quark mass, or the nucleon mass reflects
the characteristics of chiral symmetry restoration, and will
influence the pentaquark decay process $\Theta^+\rightarrow KN$.
In this letter, we investigate the effect of chiral phase
transition on the pentaquark $\Theta^+$ mass and width at finite
temperature and density at one-loop level of $N\Theta^+ K$
coupling. If the mass and width shifts induced by the chiral
symmetry restoration are sensitive to the pentaquark parity, it
may help us to solve the puzzle of $\Theta^+$ parity.

We proceed as follows. In Section 2, we calculate the self-energy
of $\Theta^+$ at finite temperature and density at one-loop level
of pseudovector and pseudoscalar $N\Theta^+ K$ couplings with
positive and negative $\Theta^+$ parity, and obtain the $\Theta^+$
mass shift and width shift in the medium. The medium dependence of
the nucleon mass is determined through the mean field gap equation
of the NJL model\cite{njl} which is one of the models to see
directly how the dynamical mechanisms of chiral symmetry breaking
and restoration operate. In Section 3, we show the numerical
calculations, analyze the contribution of chiral symmetry
restoration to the mass and width shifts, and discuss the parity
dependence of the results. Finally, we give our summary.

\section{Formulas}
We introduce the effective Lagrangians for the pseudovector and
pseudoscalar  $N\Theta^+ K$ couplings\cite{nam,kim},
\begin{eqnarray}
\label{lag} {\cal L}_{PV} &=&-\frac{g_A^\star}{2 f_\pi}
\bar\Theta^+
\gamma_\mu\gamma_5\partial^\mu K^+ n \ ,\nonumber\\
{\cal L}_{PS} &=&ig \bar\Theta^+ \gamma_5 K^+ n \ .
\end{eqnarray}
Here, the positive parity of $\Theta^+$ is assumed. The effective
lagrangians with assuming negative parity of $\Theta^+$ can be
obtained by removing $ i\gamma_5$ in the vertexes. The
pseudovector and pseudoscalar coupling constants $g_A^\star$ and
$g$ are fixed to reproduce the mass $M_\Theta=1540 MeV$ and decay
width $\Gamma_{\Theta^+}=15$ MeV at zero temperature and zero
density\cite{nam}. Through the calculation at tree level one has
$g_A^\star=0.28$ and $g=3.8$ for positive parity and
$g_A^\star=0.16$ and $g=0.53$ for negative parity\cite{nam}.

\begin{figure}[h]
\centering
\includegraphics[totalheight=0.6in]{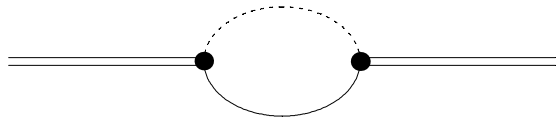}
\caption{The lowest order $\Theta^+$ self-energy. The solid and
dashed line represent nucleon and kaon field, respectively.}
\label{fig1}
\end{figure}

\begin{figure}[h]
\centering
\includegraphics[totalheight=1.5in]{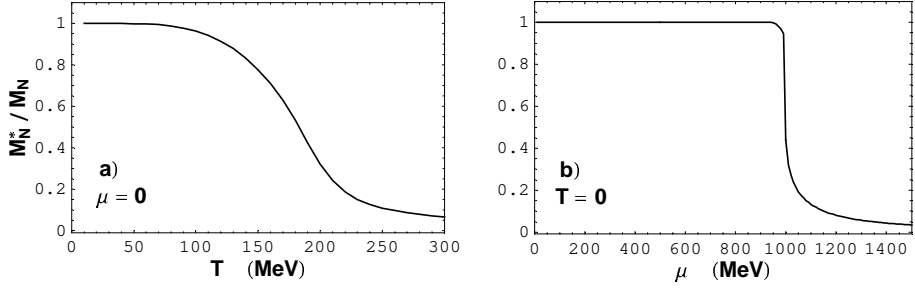}
\caption{The effective nucleon mass scaled by its value in the
vacuum as a function of temperature at zero chemical potential (a)
and a function of chemical potential at zero temperature (b).}
\label{fig2}
\end{figure}

We calculate the in-medium $\Theta^+$ self-energy by perturbation
method above mean field. To the lowest order, it is shown in
Fig.\ref{fig1}. The propagators of nucleon and kaon at mean field
level read,
\begin{eqnarray}
\label{gmf}
G_N(p) &=& \frac{i}{p'\!\!\!\!\!/-M_N}\ ,\nonumber\\
G_K(p) &=& \frac{i}{p^2-M_K^2}\ ,
\end{eqnarray}
where the four-momentum $p'$ is defined as $p'=\{p_0+\mu,{\bf
p}\}$ with baryon chemical potential $\mu$. The mechanism of
chiral symmetry restoration at finite temperature and density is
embedded in our calculation through the effective nucleon mass
$M_N$. Since the calculation of $M_N$ is nonperturbative, it is
difficult to calculate it directly with QCD, one has to use
models. While the quantitative result depends on the models used,
the qualitative temperature and density behavior is not sensitive
to the details of different chiral models\cite{li}. A simple model
to describe chiral symmetry breaking in vacuum and symmetry
restoration in medium is the NJL model\cite{njl}. Within this
model, on can obtain the hadronic mass spectrum and the static
properties of mesons remarkably well. In particular, one can
recover the Goldstone mode, and some important low-energy
properties of current algebra such as the Goldberger-Treiman and
GellMann-Oakes-Renner relations\cite{njl}. In mean field
approximation of NJL, the effective nucleon mass can be determined
through the gap equation\cite{njl,zhuang},
\begin{eqnarray}
\label{gap}
&& M_N = 3 m_q\ ,\nonumber\\
&& 1-{N_cN_fG\over \pi^2}\int_0^\Lambda dp {p^2\over
E_q}\left(\tanh{E_q+\mu/3\over 2T}+\tanh{E_q-\mu/3\over 2T}\right)
= {m_0\over m_q}\ ,
\end{eqnarray}
where $E_q=\sqrt{{\bf p}^2+m_q^2}$ is the constituent quark
energy, the current quark mass $m_0$, the color and flavor degrees
of freedom $N_c$ and $N_f$, the coupling constant $G$ and the
momentum cutoff $\Lambda$ are chosen to fit the nucleon and pion
properties in the vacuum\cite{njl,zhuang}. The numerical results
of temperature and chemical potential dependent nucleon mass are
shown in Fig.\ref{fig2}. The effective nucleon mass drops down
continuously with increasing temperature and approaches finally
three times the current quark mass $m_0$. Very different from the
temperature effect, the nucleon mass jumps down at a critical
chemical potential $\mu_c$, which means a first-order chiral phase
transition in high baryon density region. These different
temperature and density effects will be reflected in the
$\Theta^+$ mass and width shifts. Considering the near
cancellation of attractive scalar and repulsive vector potential,
the $K^+$ mass increases slightly with temperature and
density\cite{li}. To simplify the calculation we take it as a
constant $M_K=494$ MeV in the following.

%%%%%%%%%%%%%%%%%%%%%%%%%%%%%%%%%%%%%%%%%%%%%%%%%%%%%%%%%%%%%%%%%%%%%%%
%%%%%%%%%%%%%%%%%%%%%%%%%%%%%%%%%%%%%%%%%%%%%%%%%%%%%%%%%%%%%%%%%%%%%%%

%%%%%%%%%%%%%%%%%%%%%%%%%%%%%%%%%%%%%%%%%%%%%%%%%%%%%%%%%%%%%%%%%%%%%%%
%%%%%%%%%%%%%%%%%%%%%%%%%%%%%%%%%%%%%%%%%%%%%%%%%%%%%%%%%%%%%%%%%%%%%%%

The $\Theta^+$ self-energy can be separated into a scalar and a
vector part,
\begin{equation}
\label{separation}
\Sigma(p') = \Sigma_s(p')+\Sigma_\mu
(p')\gamma^\mu\ ,
\end{equation}
with which the propagator of $\Theta^+$ reads
\begin{equation}
\label{gtheta} G_\Theta = {i\over p'^\mu\gamma_\mu -m_\Theta
-\Sigma}= i
\frac{(p'^\mu-\Sigma^\mu)\gamma_\mu+(m_\Theta+\Sigma_s)}{(p'^\mu-\Sigma^\mu)(p'_\mu-\Sigma_\mu)
-(m_\Theta+\Sigma_s)^2}\ ,
\end{equation}
where $m_\Theta$ is the $\Theta^+$ mass in vacuum. The $\Theta^+$
complex mass ${\cal M}_\Theta = M_\Theta -i{\Gamma\over 2}$ in the
medium can be determined by the pole equation of the propagator,
\begin{equation}
\label{cmass}
\left[(p'^\mu-\Sigma^\mu)(p'_\mu-\Sigma_\mu)
-(m_\Theta+\Sigma_s)^2\right]\Bigg|_{p'_0=\sqrt{{\cal
M}_\Theta^2+{\bf p}^2}} = 0\ .
\end{equation}
From this equation, one can obtain the temperature, chemical
potential, and momentum dependence of the $\Theta^+$ mass and
width.

In the rest frame of $\Theta^+$, ${\bf p}=0$ and ${\bf \Sigma}=
0$, the calculation can be done more easily. Since the width of
$\Theta^+$ is very small compared to its mass(we assume that this
statement holds both in vacuum and in medium), the
Eq.(\ref{cmass}) can be separated into two uncoupled equations.
The pentaquark mass $M_\Theta$ at finite temperature and density
is calculated through the gap equation
\begin{eqnarray}
\label{mass}
M_\Theta = m_\Theta
+Re\left(\Sigma_0(M_\Theta)+\Sigma_s(M_\Theta)\right)\ ,
\end{eqnarray}
and the medium correction to the pentaquark width $\Gamma$ is
determined by
\begin{equation}
\label{width}
\Delta \Gamma_\Theta = -2
Im\left(\Sigma_0(M_\Theta)+\Sigma_s(M_\Theta)\right)\ .
\end{equation}

For the pseudovector coupling with positive parity,
\begin{eqnarray}
\label{pvp} -i\Sigma^{PV}(p) &=& -\left({g_A^\star\over
2f_\pi}\right)^2 \int\frac{d^4
k}{(2\pi)^4}(p\!\!\!/-k\!\!\!/)\frac{k'\!\!\!\!\!/-M_N}{k'^2-M_N^2}\frac{1}{(p-k)^2-M_K^2}
(p\!\!\!/-k\!\!\!/)\nonumber\\
&=& \left({g_A^\star\over
2f_\pi}\right)^2\int\frac{d^4k}{(2\pi)^4}{M_N(p-k)^2-2k'\cdot(p-k)
p'\!\!\!\!\!/+(p'^2-k'^2)k'\!\!\!\!\!/ \over
(k'^2-M_N^2)((p-k)^2-M_K^2)}\ .
\end{eqnarray}
where $k$ is the loop 4-momentum carried by the nucleon. Taking
the transformations $\int\frac{dk_0}{(2\pi)}\rightarrow iT\sum_m\
$ and $k_0\rightarrow i\omega_m$ in imaginary time formulism of
finite temperature field theory, we can obtain the explicit
expression of $\Theta^+$ self-energy at finite temperature and
density, where $\omega_m=(2m+1)\pi T$ with $m=0,\pm1,\pm2,\ldots$
is the fermion frequency. After the summation over the nucleon
frequency one derives the real and imaginary parts of the
in-medium self-energy,
\begin{eqnarray}
\label{repvp} Re\Sigma^{PV}_s(M_\Theta) &=& -{1\over
2}\left({g_A^\star\over
2f_\pi}\right)^2\int\frac{d^3\bf{k}}{(2\pi)^3}{M_N\over
E_N}\Bigg[\left(1+{M_K^2\over F_-(N,K)}\right) f_f^-
+\left(1+{M_K^2\over F_+(N,K)}\right)f_f^+\nonumber\\
&-&  M_K^2{E_N\over E_K}\left({1\over F_+(K,N)}+{1\over
F_-(K,N)}\right)f_b\Bigg]\ ,\nonumber\\
Re\Sigma^{PV}_0(M_\Theta) &=& {1\over 2}\left({g_A^\star\over
2f_\pi}\right)^2\int\frac{d^3\bf{k}}{(2\pi)^3}\Bigg[\left(1+{E_N
M_K^2+2M_\Theta{\bf k}^2\over E_N F_-(N,K)}\right)f_f^-
-\left(1+{E_N M_K^2-2M_\Theta{\bf k}^2\over E_N F_+(N,K)}\right)f_f^+\nonumber\\
&-&\left({M_\Theta(E_K^2+{\bf k}^2)\over
E_KE_N}\left({M_\Theta-E_N\over F_-(N,K)}-{M_\Theta+E_N\over
F_+(N,K)}\right) - {E_KM_K^2\over E_N}\left({1\over
F_-(N,K)}-{1\over F_+(N,K) }\right)\right)f_b\Bigg]\ ,
\end{eqnarray}
and
\begin{eqnarray}
\label{impvp} Im\Sigma^{PV}_s(M_\Theta) &=& {\pi M_N M_K^2 \over
4}\left({g_A^\star\over 2f_\pi}\right)^2\int {d^3{\bf k}\over
(2\pi)^3}{1\over E_K E_N} \Big[\left(\delta_-(K,N)
-\delta_+(K,N)\right)f_f^- + \delta_+(N,K)f_f^+\nonumber\\
&-&\left(\delta_+(K,N) +\delta_-(K,N)
-\delta_+(N,K)\right)f_b\Big]\
,\nonumber\\
Im\Sigma^{PV}_0(M_\Theta) &=& -{\pi \over 4}\left({g_A^\star\over
2f_\pi}\right)^2\int {d^3{\bf k}\over (2\pi)^3}{1\over E_NE_K}
\Bigg[(E_NM_K^2+2M_\Theta{\bf k}^2)\left(\delta_-(K,N)
-\delta_+(K,N)\right)f_f^-\nonumber\\
&-&(E_NM_K^2-2M_\Theta{\bf k}^2)\delta_+(N,K) f_f^+\nonumber\\
&-&\Bigg({M_\Theta(E_K^2+{\bf k}^2)(M_\Theta-E_N)-E_K^2M_K^2\over
E_K}\left(\delta_-(K,N)
-\delta_+(K,N)\right)\nonumber\\
&-&{M_\Theta(E_K^2+{\bf k}^2)(M_\Theta+E_N)-E_K^2M_K^2\over E_K}
 \delta_+(N,K)\Bigg)f_b\Bigg]\ ,
\end{eqnarray}
with particle energies
\begin{eqnarray}
\label{energy}
E_N^2&=&{\bf{k}^2}+M_N^2\ ,\\
E_K^2&=&{\bf{k}^2}+M_K^2\ ,
\end{eqnarray}
the Fermi-Dirac and Bose-Einstein distributions
\begin{eqnarray}\label{distri}
f_f^\pm &=&\frac{1}{e^{\left(E_N\pm\mu\right)/T}+1}\ ,\\
f_b &=&\frac{1}{e^{E_K/T}-1}\ ,
\end{eqnarray}
and the functions $F_\pm(X,Y)$ and $\delta_\pm(X,Y)$ defined by
\begin{eqnarray}\label{fun}
F_\pm(X,Y) &=& \left(M_\Theta\pm E_X\right)^2 - E_Y^2\ ,\nonumber\\
\delta_\pm(X,Y) &=& \delta\left(M_\Theta\pm E_X - E_Y\right)\ .
\end{eqnarray}

For the pseudoscalar coupling with positive parity,
\begin{eqnarray}
Re\Sigma^{PS}_s(M_\Theta) &=&-{g^2M_N\over
2}\int\frac{d^3\bf{k}}{(2\pi)^3}\left[{1\over E_N}\left({1\over
F_-(N,K) }f_f^-+{1\over F_+(N,K) }f_f^+\right)- {1\over
E_K}\left({1\over F_-(K,N)}+{1\over F_+(K,N)
}\right)f_b\right]\ ,\nonumber\\
Re\Sigma^{PS}_0(M_\Theta) &=& {g^2\over
2}\int\frac{d^3\bf{k}}{(2\pi)^3}\left[{1\over F_-(N,K) }f_f^- -
{1\over F_+(N,K) }f_f^+ -{1\over E_K}\left({M_\Theta-E_K\over
F_-(K,N)}+{M_\Theta+E_K\over F_+(K,N)}\right)f_b\right]\ ,
\end{eqnarray}
and
\begin{eqnarray}
Im\Sigma^{PS}_s(M_\Theta) &=& {\pi g^2M_N\over 4}\int {d^3{\bf
k}\over (2\pi)^3}{1\over E_K E_N} \Big[\left(\delta_-(K,N)
-\delta_+(K,N)\right)f_f^- + \delta_+(N,K)f_f^+\nonumber\\
&-&\left(\delta_+(K,N) +\delta_-(K,N)
-\delta_+(N,K)\right)f_b\Big]\
,\nonumber\\
Im\Sigma^{PS}_0(M_\Theta) &=& -{\pi g^2\over 4}\int {d^3{\bf
k}\over (2\pi)^3}{1\over E_K}\Bigg[\left(\delta_-(K,N)
-\delta_+(K,N)\right)f_f^- -\delta_+(N,K) f_f^+\nonumber\\
&-&\left({M_\Theta-E_K\over E_N}\left(\delta_-(K,N)
-\delta_+(N,K)\right)+{M_\Theta+E_K\over E_N}
\delta_+(K,N)\right)f_b\Bigg]\ .
\end{eqnarray}

For the couplings with negative $\Theta^+$ parity, the only
difference is to change the sign of the corresponding scalar self
energy, $\Sigma_s^{PV} \rightarrow -\Sigma_s^{PV}$ and
$\Sigma_s^{PS} \rightarrow -\Sigma_s^{PS}$. We will see in the
following that this change in sign leads to a partial cancellation
between $\Sigma_0$ and $\Sigma_s$ in determining the in-medium
$\Theta^+$ mass and width, and results in small mass and width
shifts for negative-parity $\Theta^+$.

\section{Numerical results}
With the formulas given in last section, we now calculate
numerically the $\Theta^+$ mass shift $\Delta M_\Theta (T,\mu) =
M_\Theta(T,\mu) - m_\Theta$ and the width shift $\Delta
\Gamma(T,\mu)$ at finite temperature and density for the
pseudovector (PV) and pseudoscalar (PS) couplings with positive
and negative $\Theta^+$ parity. We call these four couplings in
the following $PV^+, PV^-, PS^+, PS^-$, respectively.

\begin{figure}[h]
\centering
\includegraphics[totalheight=3.0in]{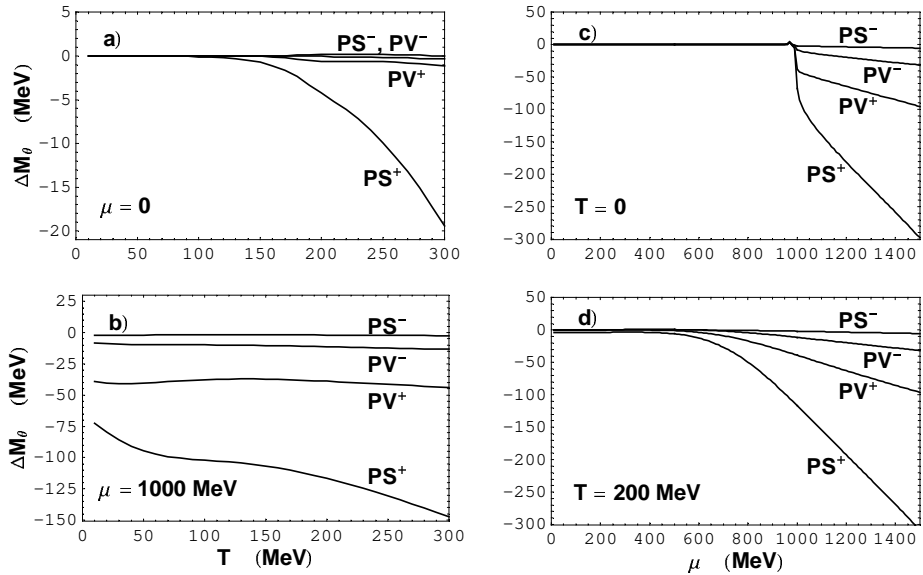}
\caption{The temperature dependence at fixed chemical potentials
$\mu =0$ (a) and $\mu=1000 MeV$ (b) and the chemical potential
dependence at fixed temperatures $T=0$ (c) and $T=200 MeV$ (d) of
the $\Theta^+$ mass shift for four different couplings.}
\label{fig3}
\end{figure}

\begin{figure}[h]
\centering
\includegraphics[totalheight=2.8in]{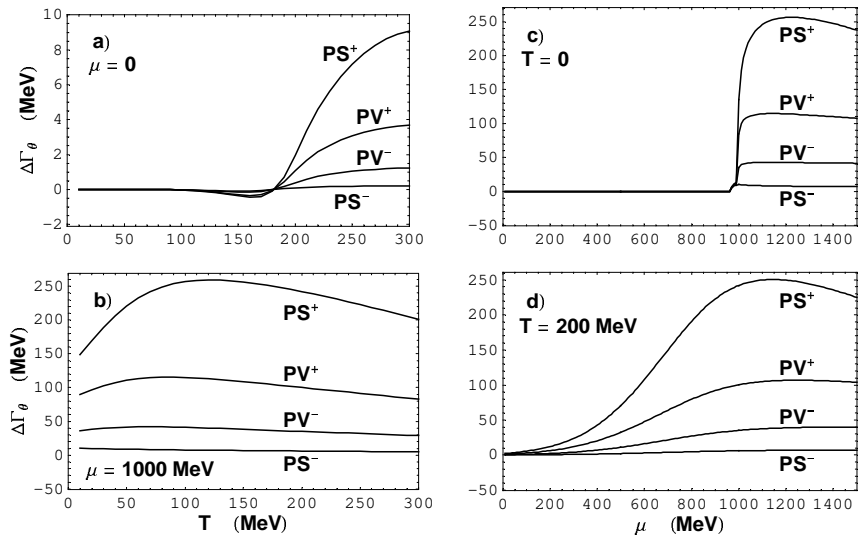}
\caption{The temperature dependence at fixed chemical potentials
$\mu=0$ (a) and $\mu=1000 MeV$ (b) and the chemical potential
dependence at fixed temperatures $T=0$ (c) and $T=200 MeV$ (c) of
the $\Theta^+$ width shift for four different couplings.}
\label{fig4}
\end{figure}

We first consider the mass shift. From its temperature dependence
at fixed chemical potentials (Figs.\ref{fig3}a and b) and chemical
potential dependence at fixed temperatures (Figs.\ref{fig3}c and
d), it has the following properties:\\
1)The $\Theta^+$ becomes light in the medium, like most of the
hadrons such as nucleon, $\rho$ meson and $\sigma$ meson
characterized by chiral symmetry. While for the couplings $PS^-,
PV^-$ and $PV^+$ the mass shift by pure temperature effect is
rather small, see Fig.\ref{fig3}a, it becomes remarkable in the
case with high baryon density and high temperature, see
Figs.\ref{fig3}b-d. \\
2) The temperature and density dependence of the mass shift is
controlled by the chiral properties. In the case of pure
temperature effect (Fig.\ref{fig3}a) and the case of high
temperature and high density effect (Figs.\ref{fig3}b and d), the
continuous chiral phase transition, shown in Fig.\ref{fig2}a for
$\mu = 0$, results in a smooth mass shift. In the case of high
density but low temperature, the mass shift is zero in the chiral
breaking phase at $\mu < \mu_c$, changes suddenly at the phase
transition with $\mu = \mu_c$, and decreases rapidly in the chiral
restoration phase with $\mu
> \mu_c$, see Fig.\ref{fig4}c. This behavior reflects the
properties of first-order chiral phase
transition shown in Fig.\ref{fig2}b.\\
3) The degree of the $\Theta^+$ mass change depends strongly on
its parity. In any case of temperature and density, the mass shift
of positive-parity $\Theta^+$ is much larger than that of
negative-parity one. For instance, at $T=200$ MeV and $\mu = 1000$
MeV, the mass shifts for the couplings $PS^+, PV^+, PV^-$ and
$PS^-$ are, respectively, $-115, -40, -10$ and $-2$ MeV.

%%%%%%%%%%%%%%%%%%%%%%%%%%%%%%%%%%%%%%%%%%%%%%%%%%%%%%%%%%%%%%%%%%%%%%%
%%%%%%%%%%%%%%%%%%%%%%%%%%%%%%%%%%%%%%%%%%%%%%%%%%%%%%%%%%%%%%%%%%%%%%%

We turn to the discussion of the $\Theta^+$ width shift. It is
indicated as a function of temperature at fixed chemical
potentials (Figs.\ref{fig4}a and b) and a function of chemical
potential at fixed temperatures (Fig.\ref{fig4}c and d) for the
four different couplings.
Related to the properties of the mass shift, the width has the characteristics:\\
1) Like most of the hadrons in medium, the suppressed mass leads
to $\Theta^+$ broadening at finite temperature and density. The
pentaquark will become more and more unstable with increasing
temperature and density, and easy to
decay in relativistic heavy ion collisions if it is created. \\
2) Again, the behavior of the width shift is dominated by the
chiral properties. The width increases continuously in the case of
pure temperature effect and the case of high density and high
temperature, but starts to jump up suddenly at the critical
chemical potential $\mu_c$ in the case of high density but low
temperature, resulted
from the chiral phase transition shown in Fig.\ref{fig2}.\\
3) The broadening depends also on the $\Theta^+$ parity. In any
case the broadening of positive-parity $\Theta^+$ is much larger
than that of negative-parity one.\\
4) Compared with the vacuum mass and width (1540 MeV and 15 MeV in
our calculation), the mass shift is slight, but the width shift is
extremely strong. From Fig.\ref{fig3} the maximum mass shift in
the considered temperature and density region is $20\%$ of the
vacuum value for the coupling $PS^+$ and $6\%$ for the coupling
$PV^+$. However, from the Fig.\ref{fig4} the maximum width shift
is 17 times the vacuum value for $PS^+$ and 7 times for $PV^+$!

%%%%%%%%%%%%%%%%%%%%%%%%%%%%%%%%%%%%%%%%%%%%%%%%%%%%%%%%%%%%%%%%%%%%%%%
%%%%%%%%%%%%%%%%%%%%%%%%%%%%%%%%%%%%%%%%%%%%%%%%%%%%%%%%%%%%%%%%%%%%%%%

To study the pentaquark production in relativistic heavy ion
collisions, we estimate the $\Theta^+$ mass and width shifts at
RHIC, SPS and SIS, shown in Tab.I. We take the corresponding
temperature and baryon chemical potential at RHIC, SPS and SIS as
$(T,\mu)=(200,50), (180,300)$ and $(150,700)$ MeV, respectively.
While the mass shift can be neglected in any case, the width shift
for positive-parity $\Theta^+$ at SPS and SIS is important and
measurable. Compared with the width shift in the vacuum, it goes
up form $7\%$ at RHIC to $40\%$ at SPS and to $380\%$ at SIS for
the coupling $PV^+$, and from $23\%$ at RHIC to $70\%$ at SPS and
to $760\%$ at SIS for the coupling $PS^+$!

\begin{table}[!ht]
\label{tab1}
\begin{center}
\begin{tabular}{|c|c|c|c|c|c|c|c|c|c|c|}\hline
\hspace{14mm}  & \hspace{8mm}  & \hspace{8mm}
& \multicolumn{4}{|c|} {$\Delta M_\Theta$} & \multicolumn{4}{|c|} {$\Delta \Gamma_\Theta$} \\
\cline{4-11}   & \raisebox{1.2ex}[0pt]{T} &
\raisebox{1.3ex}[0pt]{$\mu$} & \rule[-1mm]{0mm}{5mm}{$PV^{+}$} &
{$PV^{-}$} & {$PS^{+}$} & {$PS^{-}$} & {$PV^{+}$} & {$PV^{-}$} &
{$PS^{+}$} & {$PS^{-}$} \\ \hline RHIC & 200 & 50  & -0.5 & 0.2 &
-4.0 &   0  &  1.1 &  0.7 &  3.5  & 0.2 \\ \hline SPS  & 180 & 300
& -0.6 & 0.4 & -2.8 &  0.1 &  6.4 &  2.4 & 10.5  & 0.7 \\ \hline
SIS  & 150 & 700 & -4.0 & 1.0 & -9.4 &  0.2 & 57.6 & 20.3 & 115.1
& 4.4 \\ \hline
\end{tabular}
\end{center}
\caption{The estimation of $\Theta^+$ mass and width shifts at
RHIC, SPS and SIS for four different couplings.}
 \end{table}

\begin{figure}[h]
\centering
\includegraphics[totalheight=1.5in]{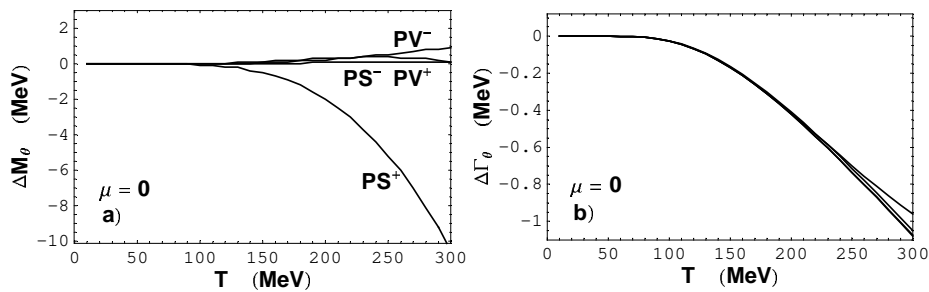}
\caption{The temperature dependence of the $\Theta^+$ mass shift
(a) and width shift (b) with a constant nucleon mass for four
different couplings.} \label{fig5}
\end{figure}

The temperature and density are introduced into our calculation
through two ways. One is the chiral symmetry restoration reflected
in the effective nucleon mass, and the other one is the loop
frequency summation in Fig.\ref{fig1}. To make sure that the
remarkable mass shift and the crucial width shift are induced
mainly by the chiral properties, we now turn off the way of chiral
phase transition and keep the nucleon mass as the vacuum value
$M_N = 940 MeV$ during the numerical calculation. The results are
shown in Fig.\ref{fig5}. Compared with the calculation with chiral
symmetry restoration, at zero chemical potential the maximum mass
shift is reduced from -20 MeV (Fig.\ref{fig3}a) to -10 MeV
(Fig.\ref{fig5}a), and the maximum width shift is strongly
suppressed from 9 MeV (Fig.\ref{fig4}a) to -1 MeV
(Fig.\ref{fig5}b). Qualitatively different from the case with
chiral symmetry restoration, the $\Theta^+$ becomes narrow in the
case without chiral symmetry restoration. It is easy to see that
the considerable mass shift, especially the extreme width shift
are originated from the mechanism of chiral phase transition.
%%%%%%%%%%%%%%%%%%%%%%%%%%%%%%%%%%%%%%%%%%%%%%%%%%%%%%%%%%%%%%%%%%%%%%%
%%%%%%%%%%%%%%%%%%%%%%%%%%%%%%%%%%%%%%%%%%%%%%%%%%%%%%%%%%%%%%%%%%%%%%%

\section{Conclusions}
We studied the temperature and density effect on the pentaquark
mass and width to the lowest order of the perturbation expansion
above chiral mean field for four different $N\Theta^+ K$
couplings. The chiral phase transition, here reflected in the
effective nucleon mass, plays an essential rule in determining the
in-medium $\Theta^+$ mass and width shifts. Like most of the
hadrons, the $\Theta^+$ becomes light and unstable in high
temperature and density region where the chiral symmetry is
restored. The degree of the mass and width shifts depends strongly
on the $\Theta^+$ parity. For positive-parity $\Theta^+$, the
maximum width shift in reasonable temperature and density region
is 17 times its vacuum value for pseudoscalar coupling, and 7
times for pseudovector coupling, while for negative-parity
$\Theta^+$, the width shift is much smaller. This parity
dependence may be helpful to determine the pentaquark parity in
relativistic heavy ion collisions. At SIS energy, the width shift
is almost 4-8 times the vacuum value for positive-parity
$\Theta^+$. When the chiral symmetry restoration is removed from
the calculation, the pure temperature and density effect resulted
from the thermal loop on the mass and width becomes rather small.
\vspace{0.3in}

\noindent {\bf \underline{Acknowledgments:}} The work is supported
in part by the Grants NSFC10135030 and G2000077407.

\end{document}